%% file: main.tex
\documentclass[a4paper, amsfonts, amssymb, amsmath, reprint, showkeys, nofootinbib, twoside]{revtex4-1}
\usepackage[english]{babel}
\usepackage[utf8]{inputenc}
\usepackage{svg}
\usepackage{xcolor}
\usepackage{comment}
\usepackage{gensymb}
\usepackage{float}

\usepackage[colorinlistoftodos, color=green!40, prependcaption]{todonotes}
\input{preamble}
\usepackage[pdftex, pdftitle={Article}, pdfauthor={Author}]{hyperref} 
\begin{document}
\title{Doping tunable charge density waves in misfit layer compounds}

\author{Hugo Le Du$^{1 \ast,\dagger}$, Ludovica Zullo$^{2 \ast,\dagger}$, Justine Cordiez$^{3 \ast}$, Robin Salvatore$^{1}$, Daniel Schmieg$^{1}$, Arindam Mukherjee$^{1}$, Giovanni Marini$^{4}$, Dominik Volavka$^{5}$, François Debontridder$^{1}$, Marie Hervé$^{1}$, Tomas Samuely$^{5}$, Shunsuke Sasaki$^{3}$, Florent Pawula$^{3}$, Etienne Janod$^{3}$, Matteo Calandra$^{4}$, Laurent Cario$^{3}$ and Tristan Cren$^{1, \dagger}$}
\affiliation{\vspace{0.3cm}
    \begin{tabular}{c} \\ $^1$Institut des Nanosciences de Paris, Paris, France \\ $^2$Institut f{\"u}r Theoretische Physik und Astrophysik and W{\"u}rzburg-Dresden Cluster of Excellence ctd.qmat, \\ Universität W{\"u}rzburg, 97074 W{\"u}rzburg, Germany \\ $^3$Institut des Matériaux de Nantes Jean Rouxel, Nantes, France \\ $^4$University of Trento, Italy \\ $^5$Centre of Low Temperature Physics, Pavol Jozef Šafárik University, Košice, Slovakia \\
    \noalign{\smallskip}
        \small $^\ast$ These authors contributed equally to this work \\
        \small $^\dagger$ Corresponding authors: hugo.ledu@insp.jussieu.fr, ludovica.zullo@uni-wuerzburg.de, tristan.cren@insp.upmc.fr
    \end{tabular}
}

\begin{abstract}
    The ability to tune charge density waves (CDWs) through external control knobs, such as doping, pressure or strain is crucial for exploring the phase diagram of two-dimensional (2D) or quasi-2D materials. Yet, controlling CDWs' critical temperature and ordering vector remains a challenge for current experimental techniques. In this work, we establish misfit layer compound heterostructures as a reliable platform to manipulate CDWs in transition metal dichalcogenides. By combining \textit{ab initio} calculations with low-temperature scanning tunneling microscopy, we show how to achieve doping-tunable control over NbSe$_2$'s CDW by chemically alloying in the rocksalt subunit. Crucially, we prove that tuning the La/Pb ratio in the misfit family (La$_x$Pb$_{1-x}$Se)$_{1.14}$(NbSe$_2$)$_2$ enables stabilization of different CDW orders, such as $2\times2$ or $3\times3$ patterns, and even coexisting phases. This work paves the way for engineering transition metal dichalcogenides with tailored charge density waves within misfit heterostructures.   
\end{abstract}

\maketitle

\input{sections/section01.tex}
\input{sections/section03.tex}
\input{sections/section04.tex}

\input{sections/section05.tex}
\input{sections/acknowledgements}

\bibliographystyle{apsrev4-1}
\bibliography{biblio}

\end{document}

%% file: preamble.tex
\usepackage{amsthm}
\usepackage{mathtools}
\usepackage{physics}
\usepackage{xcolor}
\usepackage{graphicx}
\usepackage[left=23mm,right=13mm,top=35mm,columnsep=15pt]{geometry} 
\usepackage{adjustbox}
\usepackage{placeins}
\usepackage[T1]{fontenc}
\usepackage{lipsum}
\usepackage{csquotes}
\usepackage{gensymb}

%% file: sections/section01.tex
\section{Introduction} \label{sec:Introduction}

The rich family of transition metal dichalcogenides (TMDs) exhibits a wealth of electronic orders such as superconductivity and charge density waves (CDWs). 2H-NbSe$_2$ is an archetype of this family. It hosts both superconductivity and charge density waves at low temperatures \cite{Revolinsky,Wilson}. The compound retains these competing orderings down to the monolayer limit, where the CDW remains quasi-commensurate with a $3 \times 3$ modulation.

The doping control of superconductivity in TMDs has been well established, for instance, gated MoS$_2$ develops a superconducting dome as a function of doping \cite{MoS2_gated_1,MoS2_gated_2}. By contrast, the doping control of charge density waves remains a major challenge. While 2H-NbSe$_2$ was shown to be highly tunable through mechanical strain~\cite{Gao2018, kundu_2024}, a similar control by doping is still missing for this compound. However, in 1T-ZrSe$_2$, local CDW phases could be stabilized by charge transfer from point defects~\cite{Orsted}, but this doping tunability remains spatially confined rather than uniform across the entire sample.

Two key challenges still hinder the design of doping-tunable CDWs. The first is achieving a strong and homogeneous doping at the scale of the entire sample in order to stabilize a particular CDW ground state. The second is related to common doping strategies. Field-effect gating~\cite{Das}, ionic liquid gating, chemical substitution within the TMD layer~\cite{Liu2}, and ion implantation~\cite{Li2} are limited by breakdown voltages, instability, or even sample damage. Typical doping levels remain capped near 0.14~electrons per transition metal site (or $\sim 1.4 \times 10^{14}$ e$^-$/cm$^2$).

\begin{figure*}[ht]
        \includegraphics[scale=0.35]{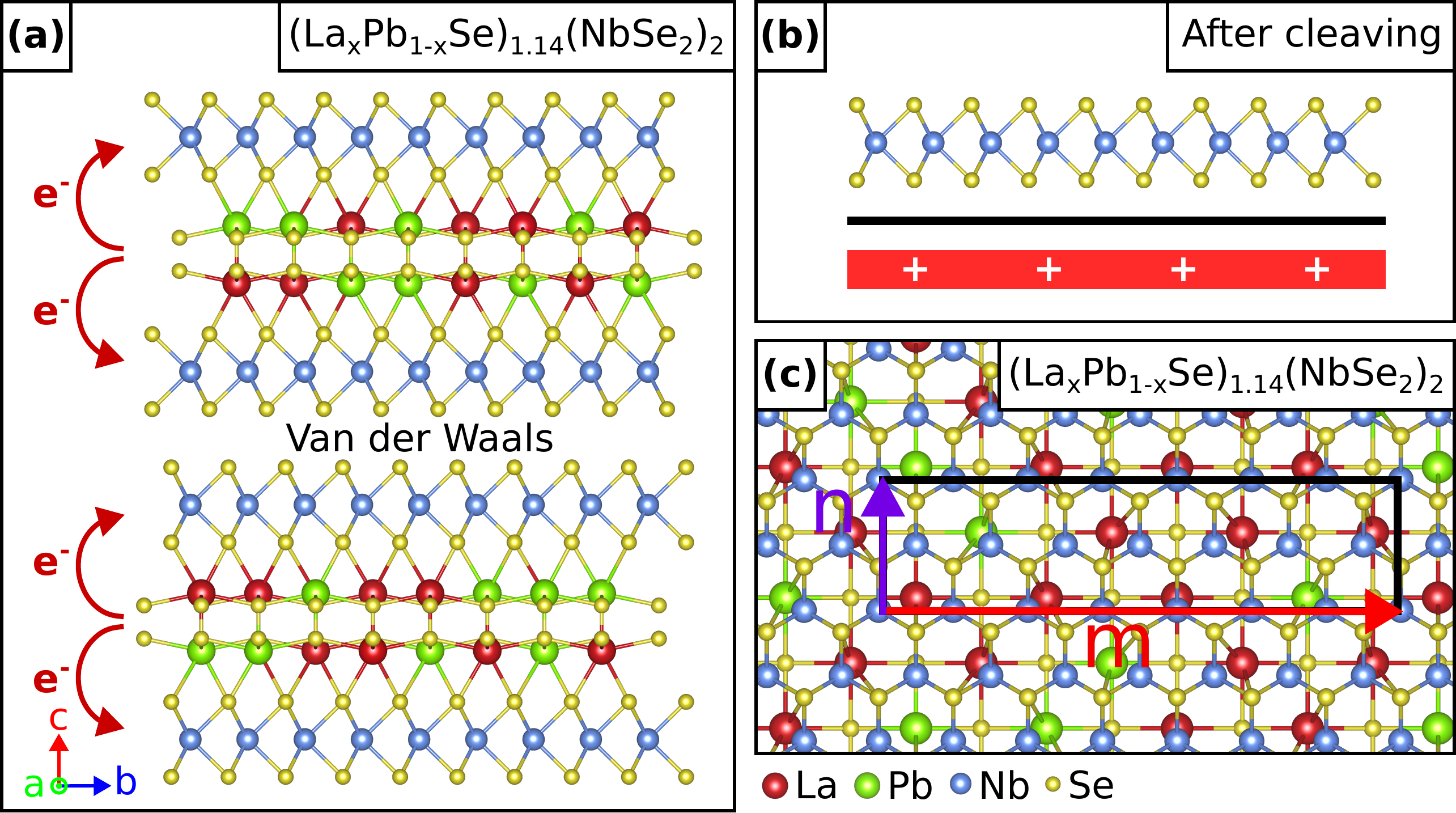} 
        \caption{a) Side-view of (La$_x$Pb$_{1-x}$Se)$_{1.14}$(NbSe$_2$)$_2$. As an example, the compound with $x=0.5$ is shown. b) The field-effect modeling scheme for the surface of misfit layer compounds. In this scheme, MLC surfaces can be modeled in a single layer field effect transistor setup (1L FET) \cite{Brumme2015,Sohier2017} by replacing the rocksalt with a positively charged gate, a negatively charged monolayer TMD then corresponds to the misfit termination layer \cite{Zullo}. To prevent the atoms to displace towards the charged plates, a positive potential barrier is added (black line). \textbf{c)} Top-view of (La$_x$Pb$_{1-x}$Se)$_{1.14}$(NbSe$_2$)$_2$. The approximate unit cell is outlined in black. The incommensurate axis, denoted m, is shown in red, while the commensurate axis n is highlighted in purple. }
      \label{fig:Figure1}
    \end{figure*}

An alternative approach for reaching heavily doped TMD layers leverages misfit layer compounds (MLCs). These are naturally formed heterostructures composed of alternating TMD and rock-salt (RS) monochalcogenide layers (Fig.~\ref{fig:Figure1}a)~\cite{Rouxel}. In these systems, strong charge transfer occurs from the RS layers to the TMDs, enabling electron doping far beyond that achievable by electrochemical means. For example, in (LaSe)$_{1.14}$(NbSe$_2$)$_2$, the NbSe$_2$ layers behave as monolayers with a Fermi level shifted by approximately 300~meV, which corresponds to a charge transfer of about 0.6 electrons per Nb atom (or $\sim 6 \times 10^{14}$ e$^-$/cm$^2$)~\cite{Leriche}.

This charge transfer in misfits is governed by two factors \cite{Zullo}. First, the work function (W) difference between TMDs and rock-salts, W(RS) $<$ W(TMD), is such that there is a preferential direction for electronic charge transfer from the RS layers to the TMDs. Second, the valence of the RS constituents is an efficient doping knob. For example, lanthanum (La$^{3+}$) donates excess charge, whereas replacing La with lead (Pb$^{2+}$) suppresses charge transfer. First-principles calculations have shown that partial substitution of La by Pb in (La$_x$Pb$_{1-x}$Se)$_{1.18}$(TiSe$_2$)$_2$ enables continuous tuning of the doping level~\cite{Zullo}.

In this work, we establish misfit layer compounds as a reliable platform to stabilize and tune charge density waves in 2D transition metal dichalcogenides. We focus on the misfit compound (La$_x$Pb$_{1-x}$Se)$_{1.14}$(NbSe$_2$)$_2$. In the chemical formula, $x$ and $1-x$ correspond to the concentrations of La and Pb atoms, respectively, in the cell (Fig.~\ref{fig:Figure1}). Iono-covalent bonding between NbSe$_2$ and (La$_x$Pb$_{1-x}$Se) layers results in stronger interfacial coupling, while adjacent NbSe$_2$ bilayers are bound by van der Waals interactions. Consequently, mechanical exfoliation typically occurs between NbSe$_2$ bilayers, yielding cleaved surfaces suitable for local probes that expose a NbSe$_2$ monolayer atop a (La$_x$Pb$_{1-x}$Se) monolayer. 

Our quest stems from a simple, yet insightful observation. Based on our DFT+ARPES study \cite{Leriche, le_du_2026}, it is possible to increase the charge transfer to NbSe$_2$ by increasing the percentage of La in the (La$_x$Pb$_{1-x}$Se)$_{1.14}$(NbSe$_2$)$_2$ samples. There are two limit cases to consider: PbSe does not donate any electrons to the transition metal dichalcogenide; thus, the surface composed only of Pb ($x=0$) can be viewed as an undoped NbSe$_2$ monolayer. In contrast, LaSe ($x=1$) leads to the maximum achievable doping concentration for this family, as each La transfers one electron to the TMD layers. Therefore, tuning the Pb/La ratio $x$ enables control of the doping, with a typical charge transfer of 0.57$x$ electrons per niobium atom in the TMD layers \cite{le_du_2026}.

Undoped monolayer NbSe$_2$ has been shown to exhibit a $3\times3$ CDW modulation. Conversely, our findings, derived from Raman spectroscopy and DFT calculations, reveal that the La misfit compound ($x=1$), corresponding to a large rigid doping of $\approx0.6$ electrons per Nb atom in NbSe$_2$ (or $\sim 6 \times 10^{14}$ e$^-$/cm$^2$), does not exhibit any CDW signature \cite{Zullo2}.

These observations naturally raise the central question: how does the CDW evolve at intermediate doping levels? Is it possible to stabilize different CDWs in NbSe$_2$ by doping via Pb-La substitution?

Our theoretical modeling based on density functional theory (DFT) calculations predicts the evolution of both the charge transfer and the associated CDW states as a function of doping. We combine these predictions with experimental studies on (La$_x$Pb$_{1-x}$Se)$_{1.14}$(NbSe$_2$)$_2$ single crystals. We use low-temperature scanning tunneling microscopy (STM), which is strictly surface-sensitive and measures only the top NbSe$_2$ layer. We use STM to perform quasiparticle interference (QPI) measurements, and our results provide local doping estimates which we correlate with observed CDW orders.

While the global doping levels in these systems have been previously established by angle-resolved photoemission spectroscopy (ARPES) \cite{le_du_2026}, our local measurements provide the necessary spatial resolution to link specific doping levels to the formation of different CDW phases. Our results show excellent agreement with both these previous global studies \cite{Leriche,Zullo,Zullo2, le_du_2026} and our theoretical predictions.

Our work demonstrates that misfit layer compounds provide a powerful and tunable platform for tailoring electronic phases in two-dimensional materials, offering new pathways to control quantum states with precision.

%% file: sections/section03.tex
\section{Results} \label{sec:Experimental}

    \begin{figure*}[ht] 
        \includegraphics[scale=0.35]{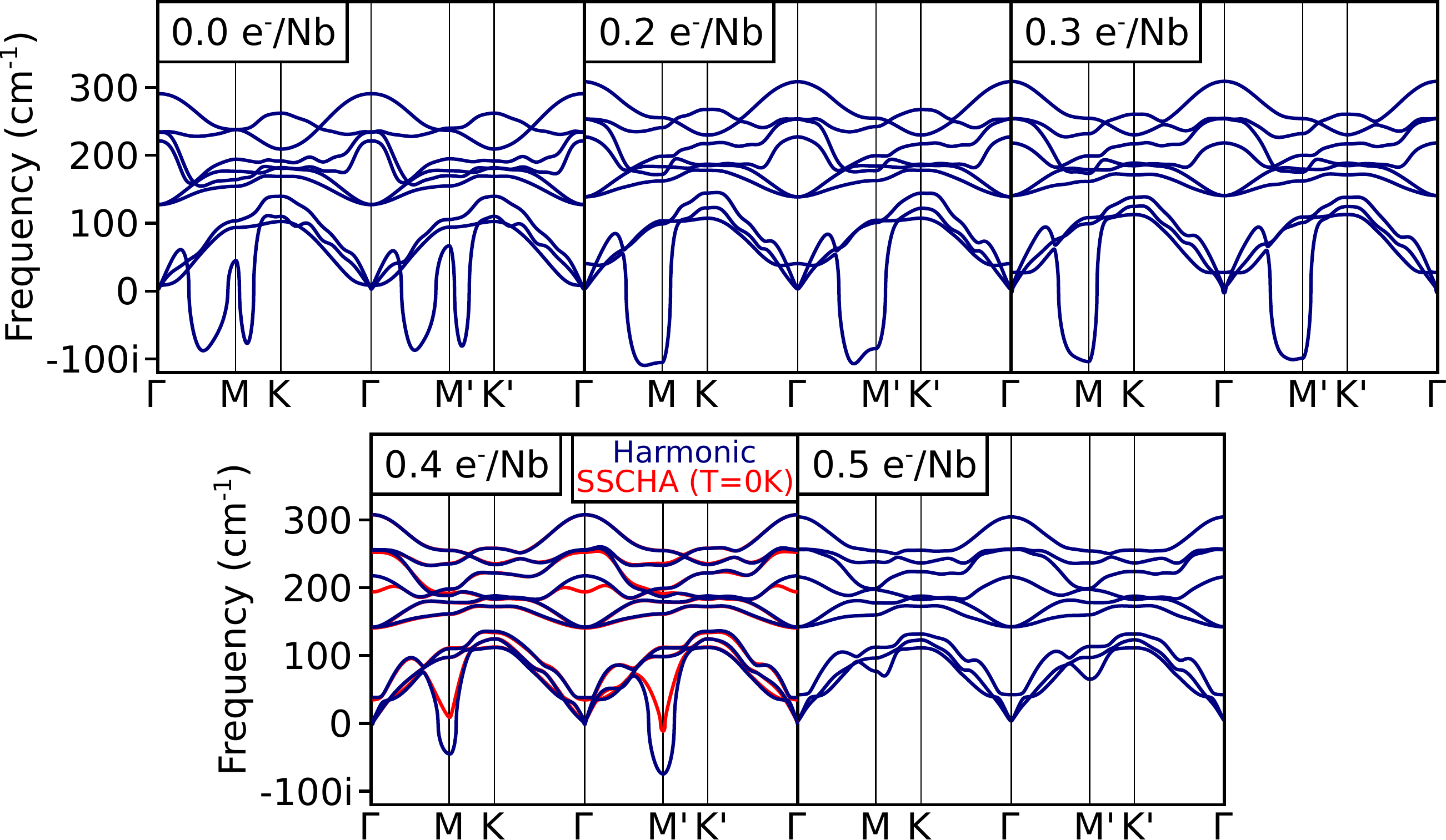} 
        \caption{Theoretical predictions for the doping evolution of the CDW of NbSe$_2$ within the misfit serie $(\text{La}_{x}\text{Pb}_{1-x}\text{Se})_{1.14}(\text{NbSe}_2)_2$. Calculated phonon dispersion along the $\Gamma$-M-K-$\Gamma$-M${'}$-K${'}$-$\Gamma$ path of a) an undoped monolayer NbSe$_2$, b)-e) a field effect doped NbSe$_2$ monolayer. Blue lines correspond to calculations in the harmonic approximation. The red line in panel d) corresponds to the anharmonic SSCHA correction. The CDW of NbSe$_2$ evolves from a $3\times3$ a) to a $2\times2$ c) and ultimately collapse at the critical doping of $0.4$ electron per Nb atom d)-e). In the intermediate doping region b) the coexistence of a $3\times3$ and a $2\times2$ orderings is observed.} 
        \label{fig:DFT}
    \end{figure*}

    \begin{figure*}[ht] 
       \includegraphics[scale=0.35]{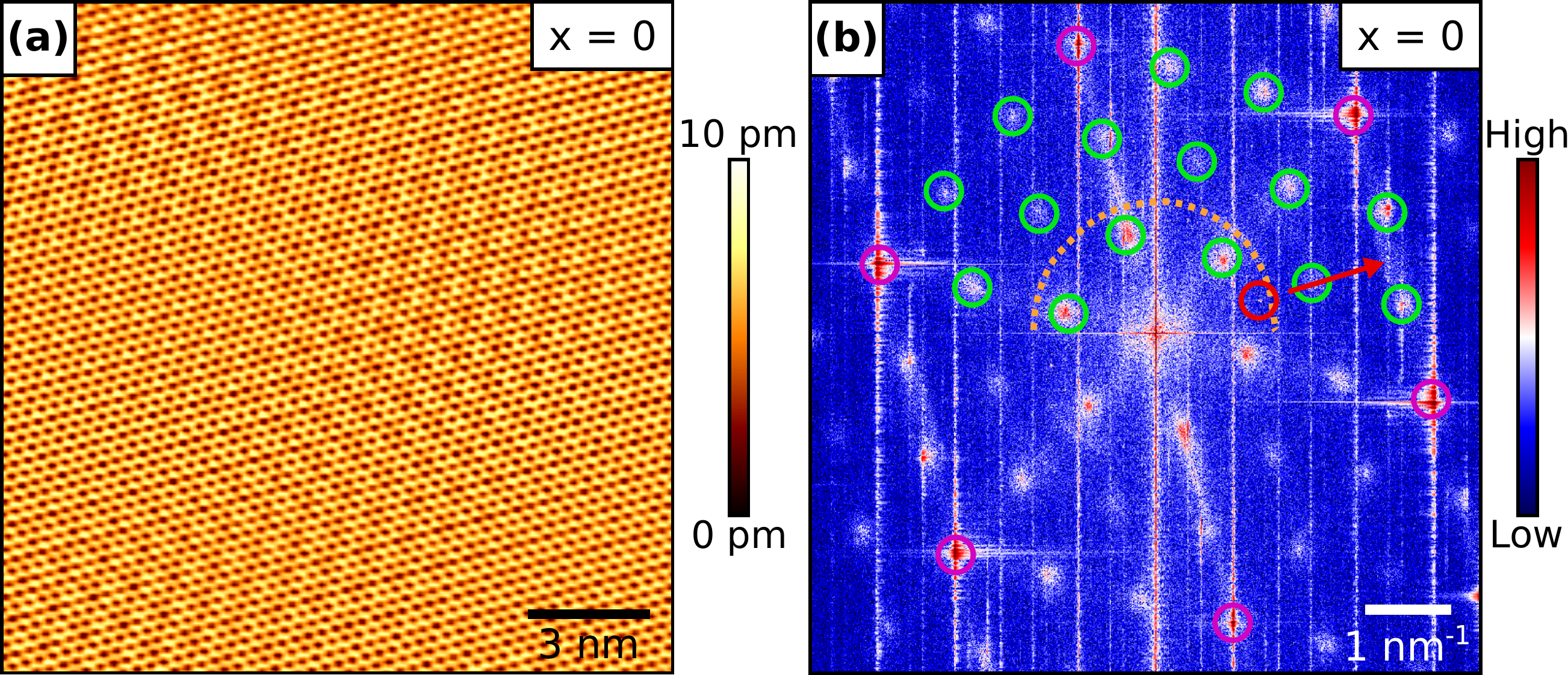} 
        \caption{a) Fourier filtered STM topography of (PbSe)$_{1.14}$(NbSe$_2$)$_2$ acquired over a $15$nm$ \times 15$nm area at a bias voltage of 20 mV and tunneling current of 200 pA, resolving a $3 \times 3$ charge density wave (CDW) modulation. b) Fast Fourier transform (FFT) of a larger $80$nm$ \times 80$nm topography. Bragg peaks are highlighted in pink, and the $3 \times 3$ CDW peaks in green. Sharp incommensurate peaks are marked in red, with their direction indicated by a red arrow. A quasiparticle interference (QPI) signal is also visible and marked in orange dashed lines.}
       \label{fig:FFT}
    \end{figure*}

To trace the CDW across different La-Pb concentrations, we first performed a predictive theoretical analysis based on density functional perturbation theory (DFPT).
To address the complexity of vibrational properties in large supercells like $(\text{La}_{x}\text{Pb}_{1-x}\text{Se})_{1.14}(\text{NbSe}_2)_2$, we employed a method for modeling the misfit as a collection of field-effect transistors by means of the 2D materials FET-setup developed in Refs. \cite{Brumme2015,Sohier2017}. This leverages the universal charge transfer mechanism in misfits, which is based on the work function difference between RS and TMDs \cite{Zullo}.
Our model allows us to effectively replicate the impact of the misfit on TMD layers by replacing the rocksalt with positively charged plates, reducing computational effort (see Fig. \ref{fig:DFT}b and Sec. A a of the SI, the optimized gate positions are listed in Tab. S1). 

We simulate cleaved misfit surfaces as a monolayer TMD within a single-gate field effect setup, as developed in Refs. \cite{Sohier2017,Brumme2015}.
The charge amount on the gate, which mimics the RS unit, is derived from the charge transfer in the full misfit calculation.
This approach was already efficiently used to estimate the misfit electronic structure in Ref. \cite{Zullo} and the CDW collapse of NbSe$_{2}$ within the $(\text{La}\text{Se})_{1.14}(\text{NbSe}_2)_2$ bulk misfit in agreement with Raman measurements~\cite{Zullo2}. 

In all the calculations we considered a quasi-hexagonal monolayer NbSe$_2$. Experimental structural study has shown that, as compared to bulk 2H-NbSe$_2$, the layers' lattice is slightly compressed along one of the in-plane directions within the misfit \cite{Leriche}. Thus, the actual space group of NbSe$_2$ is C2221, with the angle of $60$\degree~along the misfit direction compressed to $59.6$\degree. To account for this strain, we set the in-plane lattice parameters $a_{1} \neq b_{1}$ to $3.457$\AA \ and $3.437$\AA \, respectively. It should be noted that this effect is tiny and gives negligible contributions for electronic properties of the misfits \cite{Leriche} and vibrational properties of non-CDW systems \cite{Zullo2}. However, because the strain imposed by the misfit breaks the equivalence of the M and M$^{'}$ phonons, accounting for this symmetry constraint becomes crucial when considering CDW systems. A comparison with the hexagonal phonon calculations is reported in Fig. S1.

The theoretical predictions are summarized in Fig.\ref{fig:DFT} and Tab. \ref{tab:CDW}. The evolution of the CDW of NbSe$_2$ across different doping concentrations is showcased. Starting from the undoped NbSe$_2$ monolayer, which corresponds to the only-Pb case, we observe the usual $3\times3$ CDW. 
As the doping with the La concentration increases, the CDW evolves: crucially, intermediate doping calculations show the coexistence of a $3\times3$ and a $2\times2$ CDW. Finally for high doping regimes (60$\%$ to 80$\%$ La content) the system is driven towards the collapse of the CDW. Ultimately, the $100\%$ La compound shows no signature of CDW as already demonstrated in our early work \cite{Zullo2}.
These results align perfectly with the previous reasoning and will be further discussed in the following.

The blue lines in Fig. \ref{fig:DFT} correspond to harmonic phonon calculations. The harmonic approximation effectively predicts the system's tendency toward instabilities, providing information about the trends of charge density waves with doping. On the other hand, it has been previously demonstrated that the inclusion of anharmonicity hardens the phonon frequencies for an undoped NbSe$_2$ monolayer \cite{Bianco_2020}. Considering that CDW is suppressed with increased doping, we opted to conduct stochastic self-consistent harmonic approximation (SSCHA) \cite{Monacelli_2021} calculations to accurately assess the critical doping value for the breakdown of the CDW of NbSe$_2$ within the misfit (computational details and convergence tests in Sec. A of the SI and Fig. S2). In particular, we considered the compound with $50\%$ of the La content with a corresponding charge transfer of $0.4$ electrons per Nb atoms. From harmonic calculations, two imaginary phonons are predicted at M and M$^{'}$ points, suggesting a $2\times2$ CDW ordering. However, within the SSCHA correction, (red lines in Fig. \ref{fig:DFT}d) the instabilities are strongly reduced. Only a slightly unstable phonon of the order of -10i cm$^{-1}$ is reported at M$^{'}$ point, probably disappearing with small thermal fluctuations. Therefore, we conclude that $50\%$ La content is the critical value for the existence of a CDW.


Given the promising theoretical results, to confirm our predictions, we performed STM experiments.
The electronic properties of misfit layer compounds (La$_x$Pb$_{1-x}$Se)$_{1.14}$(NbSe$_2$)$_2$ demonstrate significant charge transfer from the RS layer to NbSe$_2$ layers~\cite{le_du_2026}. Although characterized by Raman spectroscopy, energy-dispersive X-ray analysis (EDX), and ARPES, local compositional inhomogeneities may be present, highlighting the necessity for a method to evaluate local doping.
A representative STM topography acquired on (PbSe)$_{1.14}$(NbSe$_2$)$_2$ is shown in Fig. \ref{fig:FFT}a (the corresponding raw, unfiltered topography is shown in Fig. S9). 
The image, recorded with upward scan direction and horizontal fast axis, resolves both the atomic lattice and a prominent $3 \times 3$ charge density wave. The corresponding Fourier transform (Fig. \ref{fig:FFT}b) clearly reveals sharp peaks arising from the lattice incommensurability. It also displays peaks associated with the CDW, consistent with the real-space $3 \times 3$ modulation similar to that observed in 2H-NbSe$_2$. This similarity is expected, as the absence of electron transfer from the RS to the NbSe$_2$ layers in (PbSe)$_{1.14}$(NbSe$_2$)$_2$ leaves the NbSe$_2$ layers undoped, analogous to 2H-NbSe$_2$. 

\begin{figure*}
        \includegraphics[scale=0.35]{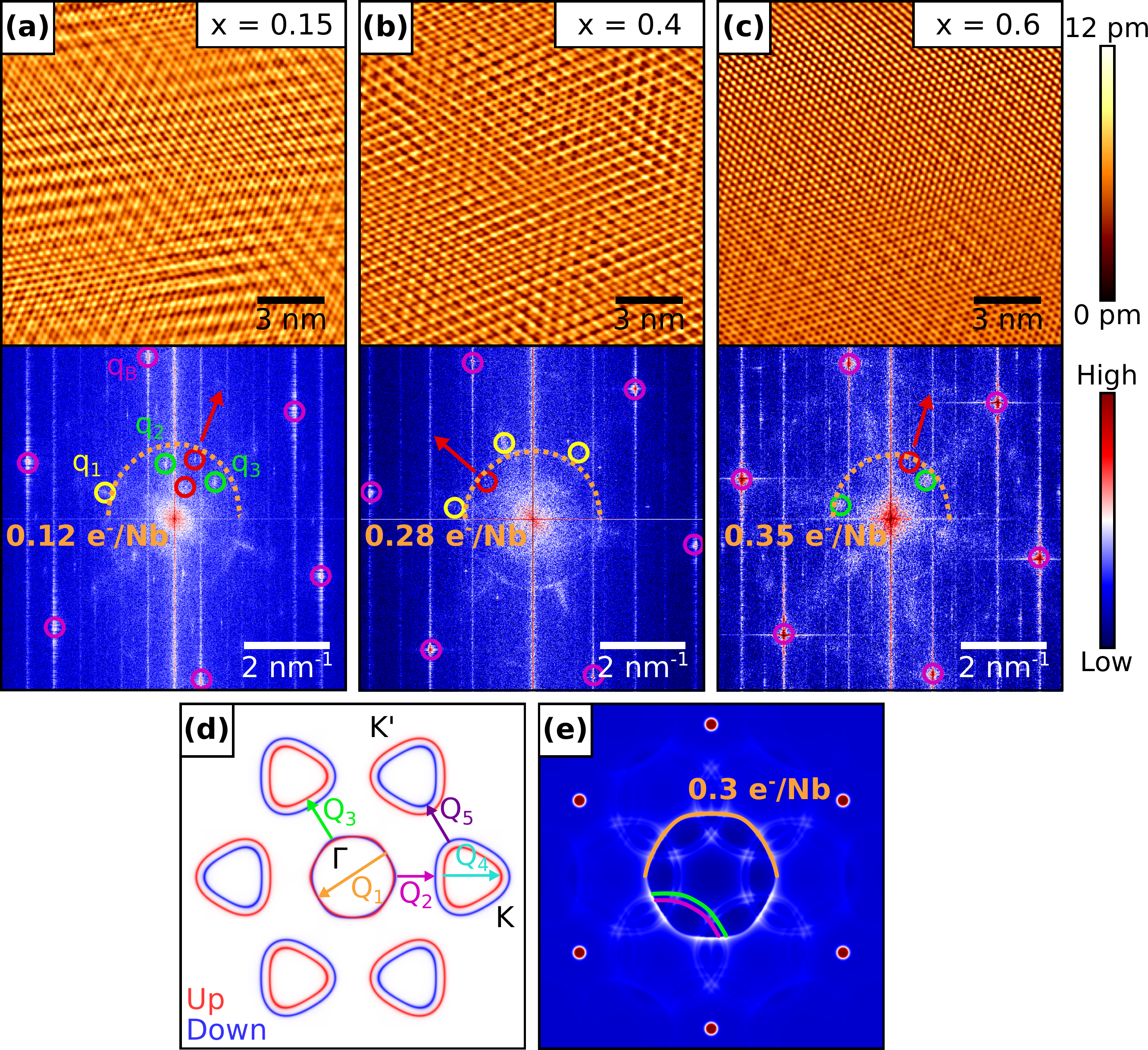} 
        \caption{a-c) Fourier filtered STM topographies of different (La$_x$Pb$_{1-x}$Se)$_{1.14}$(NbSe$_2$)$_2$ compounds acquired over a $15$nm$ \times 15$nm area at a bias voltage of 20 mV and tunneling current of 200 pA, each compound displays a different CDW pattern. For each compound, fast Fourier transforms (FFTs) of larger-area topographic maps ($80$nm$ \times 80$nm) are displayed, revealing distinct features in reciprocal space. Bragg peaks are highlighted in pink, while $3 \times 3$ and $2 \times 2$ CDW order appears in green and yellow, respectively. Sharp incommensurate peaks are marked in red, with their direction indicated by a red arrow. In addition, a quasiparticle interference (QPI) pattern is visible and highlighted in orange. The Fermi levels of the samples, compared to undoped monolayer NbSe$_2$, show systematic shifts: these shifts correlate with the La content in the RS layer. The CDW ordering also evolves with RS content. d) Fermi surface of doped NbSe$_2$ exhibiting three distinct pockets: $\Gamma$, $K$, and $K'$. Ising spin–orbit coupling induces spin-dependent band splitting, particularly evident near the $K$ and $K'$ valleys. Representative scattering vectors associated with various scattering processes are indicated in distinct colors. e) Theoretical QPI patterns, computed using the T-matrix formalism based on the electron-doped NbSe$_2$ Fermi surface, revealing distinct features corresponding to specific scattering processes. Each component of the pattern is color-coded to match its associated scattering vector, while Bragg peaks are marked in red. }
        \label{fig:FFT2}
    \end{figure*}

To investigate the evolution of the NbSe$_2$ CDW as a function of the misfit layer stoichiometry, we performed high-resolution STM imaging on five distinct (La$_x$Pb$_{1-x}$Se)$_{1.14}$(NbSe$_2$)$_2$ heterostructures with varying La content from $x=0$ to $x=0.6$ at T$=4.2$ K and V$_{\text{bias}} =20$ mV under UHV. Crystal growth is detailed in Sec. A of the SI. As summarized in Figure \ref{fig:FFT2} and Table \ref{tab:CDW}, the CDW symmetry undergoes a complex evolution with chemical composition: at low doping ($x=0.15$ and $0.3$), we observe a coexistence of $3 \times 3$ and $2 \times 2$ modulations exhibiting distinct crystallographic orientations (discussed further). At an intermediate concentration of $x=0.4$, the $3 \times 3$ phase is entirely suppressed in favor of a pure $2 \times 2$ order, while a further increase in La content ($x=0.6$) triggers a re-emergence of the pure $3 \times 3$ CDW. Notably, all observed phases remain quasi-commensurate, reminiscent of the $3 \times 3$ CDW in 2H-NbSe$_2$. The maximum coherence length $\xi$, derived from the inverse full-width at half-maximum (FWHM) of the CDW peaks in the Fourier transform, ranges from over 16 nm at low doping to below 4 nm for $x = 0.6$.

To probe the doping of each sample and correlate it with observed CDW, we exploit the circular patterns observed in all FFTs (highlighted in orange in Fig. \ref{fig:FFT}b), which arise from quasiparticle interference (QPI). We attribute these patterns to electron scattering off the periodic potential created by the incommensurate lattice. As illustrated in Fig. \ref{fig:FFT2}d, both intrapocket ($\Gamma$–$\Gamma$, $K$–$K$) and interpocket ($\Gamma$–$K$, $K$–$K'$) scattering processes contribute. Due to strong Ising spin–orbit coupling (SOC), $K$-point scattering features are spin-split, whereas $\Gamma$-point processes remain largely unaffected. 
    
The characteristic QPI vectors provide a fingerprint of the Fermi surface geometry and hence the doping level. Given that (La$_x$Pb$_{1-x}$Se)$_{1.14}$(NbSe$_2$)$_2$ is expected to behave as an electron-doped monolayer NbSe$_2$, we simulate QPI using T-matrix formalism within the rigid-band approximation, validated by density functional theory (DFT) \cite{Zullo, Zullo_thesis} and ARPES~\cite{le_du_2026} (see Sec. B, Tab. S2 and Figs. S3–S4 of the SI). An example of a computed QPI map for a doped monolayer is shown in Fig. \ref{fig:FFT2}e. The dominant $\Gamma$–$\Gamma$ scattering yields a rounded hexagonal feature, while weaker $\Gamma$–$K$ contributions (vectors $\textbf{Q}_2$ and $\textbf{Q}_3$) encode information about Ising SOC.

By comparing the size of the observed hexagonal QPI contour to the simulated maps (see Fig. S6), we extract the Fermi level shifts and estimate the doping level for each composition (Fig. \ref{fig:FFT2}a–c). This approach reveals a clear trend: doping increases with La content, consistent with electron donation from the LaSe rocksalt layer to NbSe$_2$.

\begin{table}[]
\begin{tabular}{|cccccc|}
\hline
\multicolumn{6}{|c|}{DFT} \\ \hline
\multicolumn{1}{|c|}{Doping (e$^-$/Nb)} & 0.0 & 0.2 & 0.3 & 0.4 & 0.5 \\ \hline
\multicolumn{1}{|c|}{CDW} & 
\begin{tabular}[c]{@{}c@{}}$3 \times 3$\end{tabular} & 
\begin{tabular}[c]{@{}c@{}}$3 \times 3$\\ $~~+~2 \times 2~$\end{tabular} & 
\begin{tabular}[c]{@{}c@{}} $~~~2 \times 2~~$\end{tabular} & 
-- & 
-- \\ \hline

\multicolumn{6}{|c|}{STM} \\ \hline
\multicolumn{1}{|c|}{Doping (e$^-$/Nb)} & 0.0 & 0.12 & 0.21 & 0.28 & 0.35 \\ \hline
\multicolumn{1}{|c|}{CDW}       & 
\begin{tabular}[c]{@{}c@{}}$3 \times 3$\end{tabular} & 
\begin{tabular}[c]{@{}c@{}}$3 \times 3$\\ $+~2 \times 2$\end{tabular} & 
\begin{tabular}[c]{@{}c@{}}$3 \times 3$\\ $+~2 \times 2$\end{tabular} & 
$~2 \times 2~$ & 
$~3 \times 3~$ \\ \hline
\multicolumn{1}{|c|}{$\xi_{\text{max}}^{3 \times 3}$ (nm)} & 16.1 & 12.6 & 9.2 & - & 3.3 \\ \hline
\multicolumn{1}{|c|}{$\xi_{\text{max}}^{2 \times 2}$ (nm)} & - & 12.4 & 12.1 & 10.5 & - \\ \hline
\multicolumn{1}{|c|}{La content} & 0.0 & 0.15 & 0.3 & 0.4 & 0.6 \\ \hline
\end{tabular}

\caption{Summary of DFT and STM results. CDW transitions from a pure $3\times3$ state to a coexistence of $2\times2$ and $3\times3$ modulations, then to a pure $2\times2$ phase, and ultimately leads to the suppression of CDW order. The coherence length of each sample corresponds to the maximum coherence length observed in that sample (exemples of raw FFTs for all compositions are presented in Fig. S10).}
\label{tab:CDW}
\end{table}

%% file: sections/section04.tex
\section{Discussion} \label{sec:Analysis}
    A comprehensive comparison of the STM and DFT results is summarized in Table \ref{tab:CDW}. The observed CDW evolve systematically with doping: the canonical $3 \times 3$ order weakens with increasing electron density, giving way to a $2 \times 2$ phase that ultimately vanishes at high doping.

    We begin by examining the undoped limit, where the system is expected to recover the characteristic properties of 2H-NbSe$_2$. As anticipated, the misfit compound (PbSe)$_{1.14}$(NbSe$_2$)$_2$ exhibits a well-defined $3 \times 3$ CDW, consistent with the behavior of undoped NbSe$_2$. This is consistent with our DFPT calculations showing the CDW instability at $2/3$ $\Gamma-M$ q-point (Fig. \ref{fig:DFT}a).
    
    QPI measurements for (PbSe)$_{1.14}$(NbSe$_2$)$_2$ reveal significant deviations from theoretical QPI patterns derived within the rigid band approximation (Fig.~\ref{fig:FFT}). We attribute this discrepancy to the presence of additional spectral weight in the valence band. As confirmed from our band unfolding DFT calculations in Ref. \cite{le_du_2026}, this is associated with few PbSe-derived states crossing Nb-d band of NbSe$_2$ within the misfit. These states may obscure or modify the QPI signal intrinsic to NbSe$_2$. Nevertheless, this effect does not hinder the zero charge transfer limit predicted for this compound, nor the persistence of the $3\times3$ CDW.
    
    We now turn to the low-to-intermediate doping regime, focusing on (La$_{0.15}$Pb$_{0.85}$Se)$_{1.14}$(NbSe$_2$)$_2$ and (La$_{0.3}$Pb$_{0.7}$Se)$_{1.14}$(NbSe$_2$)$_2$, where we observe the coexistence of $2 \times 2$ and $3 \times 3$ CDW phases.
    This doping regime corresponds approximately to a charge transfer of 0.09-0.15 electrons per Nb atom. In this regime, our theoretical calculations point out a possible coexistence of the two competing CDWs. Taking into account anharmonic corrections, the coexistence between the $3\times3$ CDW and the $2\times2$ CDW occurs at approximately $0.2$ electrons per Nb atom.
    
    In (La$_{0.15}$Pb$_{0.85}$Se)$_{1.14}$(NbSe$_2$)$_2$, the $2 \times 2$ modulation emerges along a single crystallographic direction, specifically, perpendicular to the incommensurate direction, as indicated by the $q_1$ orientation in Fig. \ref{fig:FFT2}. Along the other two directions, corresponding to vectors $q_2$ and $q_3$, the conventional $3 \times 3$ modulation persists.
    
    The emergence of this twofold anisotropy, which breaks the intrinsic threefold symmetry of 1H-NbSe$_2$, is likely driven by uniaxial strain. This interpretation is based on the well-known sensitivity of CDW to lattice deformations in TMDs \cite{Gao2018, chiu2025, merino2025, zhang2017}. In our system, the NbSe$_2$ lattice stretches along the commensurate direction while compressing along the incommensurate axis in order to match the lattice of the RS layer along the commensurate direction. This anisotropic deformation reduces the original threefold symmetry. Importantly, this symmetry breaking differs from a conventional nematic transition. Here, it arises from extrinsic lattice constraints imposed by the misfit strain, which defines a preferential structural axis. This structural perspective provides an alternative viewpoint to recent studies on TaS$_2$ misfits, where an electronic symmetry-breaking field from the Moiré potential was suggested as the primary driver for lifting the threefold degeneracy \cite{sajan2026moireinduced}.
    
    While strain-induced CDW transitions have previously been reported in NbSe$_2$ \cite{Gao2018, chiu2025}, strain alone cannot fully account for the observed phenomena. Similar levels of strain are present in related compounds, such as the undoped (PbSe)$_{1.14}$(NbSe$_2$)$_2$, which do not exhibit the $2 \times 2$ phase. Instead, the effectiveness of this structural coupling is significantly enhanced by the fact that the system is chemically tuned to the vicinity of the doping-induced transition between the $3 \times 3$ and $2 \times 2$ CDW phases. In this transitional regime, the electronic order becomes intrinsically fragile and highly susceptible to external perturbations. Consequently, strain may play a stabilizing role, either promoting the $2 \times 2$ CDW along $q_1$ or reinforcing the $3 \times 3$ order along $q_2$ and $q_3$, where it might otherwise be unstable.

    In the Fourier transform of topography on (La$_{0.3}$Pb$_{0.7}$Se)$_{1.14}$(NbSe$_2$)$_2$ (see Fig. S10), both $2 \times 2$ and $3 \times 3$ modulations also appear along distinct crystallographic directions. In addition, broad diffuse features are present at both $3 \times 3$ and $2 \times 2$ positions, suggesting competition or coexistence between the two orders. 

    Furthermore, in  (La$_{0.4}$Pb$_{0.6}$Se)$_{1.14}$(NbSe$_2$)$_2$ (0.28 electrons per Nb atom) we observe a pure $2 \times 2$ CDW state in all directions, consistent with our calculations for a doping of $0.3$ electrons per Nb atoms taking into account anharmonicity. For higher doping level, our precise SSCHA calculation revealed a complete collapse of any charge density wave ordering, starting from a critical doping of $0.4$ electrons per Nb atom (see red line in Fig. \ref{fig:DFT}d). This corresponds to a $50\%$ La content in the misfit structure.

    The $2 \times 2$ CDW manifests a pronounced spatial inhomogeneity, with correlation lengths $\xi$ fluctuating between 3 and 12 nm depending on both crystallographic orientation and local sampling. Such significant variations underscore a heightened sensitivity to quench disorder, suggesting that the $2 \times 2$ phase is intrinsically fragile and easily disrupted at the nanoscale. Notably, the FFTs reveal highly anisotropic peaks: while the CDW maintains a moderate degree of coherence along the direction of Bragg peaks, it exhibits a sharp increase in coherence length in the transverse direction. This anisotropy is characteristic of a ribbon-like morphology in real space, where the charge order is elongated along a preferential axis but remains strictly confined laterally.

    To summarize, our observations, supported by DFT calculations, point to a continuous transition from the $3 \times 3$ to the $2 \times 2$ CDW phase as doping increases. The ability to tune the CDW state through chemical substitution highlights the role of misfit as a platform for doping control. At the transition, the CDW becomes increasingly fragile and susceptible to external perturbations, here, strain appears to favor the stabilization of the $2 \times 2$ order over the $3 \times 3$ phase. However, doping remains the primary driving parameter in this transition.

    Finally, we discuss the high-doping regime, where our DFT calculations predict the collapse of CDW order. In previous work on (LaSe)$_{1.14}$(NbSe$_2$)$_2$, corresponding to a heavily electron-doped NbSe$_2$, a complete suppression of the CDW phase was demonstrated through first-principles calculations and polarized Raman spectroscopy~\cite{Zullo2}. Early STM measurements revealed a short-range $2 \times 2$ modulation. However, these findings are attributed to local pinning by surface impurities rather than a robust bulk phase~\cite{Leriche}.

    Here, we investigate a slightly lower doping regime corresponding to $60\%-80\%$ of La content in the sample. In this case, still a high charge transfer of $\approx$ $0.5$ electrons per Nb atom occurs. For such a doping, our theoretical calculations predict the disappearance of long-range CDW order (see Fig. \ref{fig:DFT}). Inspecting the phonon spectra, no instability is reported, as the soft phonon mode that was previously present at M and M$^{'}$ hardens. 
    
    Our STM measurements on (La$_{0.6}$Pb$_{0.4}$Se)$_{1.14}$(NbSe$_2$)$_2$ indeed show that the $2\times2$ order is not stabilized as predicted. However, a resurgence of a $3\times3$ CDW order is found, with associated Bragg peaks that appear significantly broadened compared to the sharper $2 \times 2$ and $3 \times 3$ peaks seen in the less doped compounds. Quantitatively, we extract a CDW coherence length of ~3 nm in (La$_{0.6}$Pb$_{0.4}$Se)$_{1.14}$(NbSe$_2$)$_2$, in contrast to the $\geq$ 10 nm coherence typical of other compounds exhibiting similar ordering. 
    
    This is reminiscent of prior observations in the highly doped limit, where STM detected only very short-range order, undetectable by Raman spectroscopy and absent in DFT predictions. The presence of such local modulations likely reflects disorder-induced instabilities. Similar phenomena have been reported in bulk 2H-NbSe$_2$, where structural disorder was shown to stabilize a short-range $3 \times 3$ CDW well above the critical temperature \cite{Arguello2014}.

    Together, these results point to the onset of CDW collapse at intermediate doping levels, with disorder playing a secondary role in enabling residual short-range order.

    Finally we comment on the origin of different CDW modulations in NbSe$_2$. The formation of the $3 \times 3$ CDW in NbSe$_2$ is well understood and attributed to strong electron–phonon coupling \cite{weber2013, johannes_fermi_2008}. In contrast, the emergence of a $2 \times 2$ CDW has not been observed in undoped or unstrained NbSe$_2$, and its mechanism remains less clear.

    One natural hypothesis is that the $2 \times 2$ modulation arises from Fermi surface nesting. To test this scenario, we calculated the imaginary part of the electronic susceptibility using the Fermi surface of a doped NbSe$_2$ monolayer. However, the results reveal no prominent nesting features at the $M$ point, effectively ruling out nesting as the primary driving force (see Sec. D and Figs. S7–S8 of the SI).

%% file: sections/section05.tex
\section{Conclusion} \label{sec:conclusion}
    Systematic synthesis and investigation of (La$_x$Pb$_{1-x}$Se)$_{1.14}$(NbSe$_2$)$_2$ single crystals reveals a tunable electronic landscape in NbSe$_2$ driven by chemical doping. By varying the La/Pb ratio, the Fermi level of the NbSe$_2$ layers can be modulated over a substantial range, corresponding to a shift in Fermi energy of up to $0.3$ eV in the La-rich limit. This doping-induced tuning is accompanied by a transformation of the CDW order, evolving from the canonical $3 \times 3$ CDW phase at low doping to a $2 \times 2$ state at intermediate doping, with suppression of CDW order at high doping levels. Further Raman spectroscopy measurements would offer a powerful probe of this transition, enabling distinction between bulk and surface CDW contributions and highlighting the complex interplay between CDW order and superconductivity in this system.

    More broadly, our findings underscore the potential of misfit layer compounds as a versatile platform for tuning and exploring the emergent electronic phases of TMDs. Previous theoretical studies have demonstrated superconductivity tuning in (La$_x$Pb$_{1-x}$Se)$_{1.18}$(TiSe$_2$)$_2$, our results extend this paradigm to CDW and suggest the possibility of controlling other phenomena such as the Mott transition in related 2D systems, including TaS$_2$. These results open pathways for the controlled design of quantum phases in low-dimensional materials via chemical synthesis.

%% file: sections/acknowledgements.tex
\section*{Acknowledgements} \label{sec:acknowledgements}

L.Z. gratefully acknowledges the scientific support and HPC resources provided by the Erlangen National High Performance Computing Center (NHR@FAU) of the Friedrich-Alexander-Universität Erlangen-N{\"u}rnberg (FAU) under the NHR project b294cb. NHR funding is provided by federal and Bavarian state authorities.

This work was supported by the French Agence Nationale de la Recherche through the contract ANR Misfit (ANR-21-CE30-0054), the ANR SURIKAT (ANR-23-CE30-0036) and the ANR MASCOTE (ANR-24-CE30-1342).

%% file: biblio.bib
@article{Orsted,
  author = {Ørsted, A. and Scarfato, A. and Barreteau, C. and Giannini, E. and Renner, C.},
  title = {Doping Tunable CDW Phase Transition in Bulk 1T-ZrSe2},
  journal = {Nano Letters},
  volume = {25},
  number = {4},
  pages = {1729--1735},
  year = {2025}
}

@article{Revolinsky,
  author = {Revolinsky, E. and Spiering, G. and Beerntsen, D.},
  title = {Superconductivity in the niobium-selenium system},
  journal = {Journal of Physics and Chemistry of Solids},
  volume = {26},
  pages = {1029},
  year = {1965}
}

@article{Wilson,
  author = {Wilson, J. and Di Salvo, F. and Mahajan, S.},
  title = {Charge-density waves and superlattices in the metallic layered transition metal dichalcogenides},
  journal = {Advances in Physics},
  volume = {24},
  pages = {117},
  year = {1975}
}

@article{Das,
  author = {Das, S. and Sebastian, A. and Pop, E. and others},
  title = {Transistors based on two-dimensional materials for future integrated circuits},
  journal = {Nature Electronics},
  volume = {4},
  pages = {786--799},
  year = {2021}
}

@article{Liu2,
  author = {Liu, J. and Li, B. and Li, Q.},
  title = {Two-Dimensional Doped Materials},
  journal = {Magnetochemistry},
  volume = {8},
  pages = {172},
  year = {2022}
}

@article{Li2,
  author = {Li, Ziqi and Chen, Feng},
  title = {Ion beam modification of two-dimensional materials: Characterization, properties, and applications},
  journal = {Applied Physics Reviews},
  volume = {4},
  number = {1},
  pages = {011103},
  year = {2017}
}

@article{Rouxel,
  author = {Rouxel, J. and Meerschaut, A.},
  title = {Misfit Layer Compounds (MX)n(TX2)m [M=Sn,Pb,Bi,Rare Earth; T=Transition Metal; X=S,Se; 1.08<n<1.25; m=1,2]},
  journal = {Molecular Crystals and Liquid Crystals},
  volume = {244},
  year = {1994}
}

@article{Leriche,
  author = {Leriche, Raphaël T. and Palacio‐Morales, Alexandra and Campetella, Marco and Tresca, Cesare and Sasaki, Shunsuke and Brun, Christophe and Debontridder, François and David, Pascal and Arfaoui, Imad and Šofranko, Ondrej and Samuely, Tomas and Kremer, Geoffroy and Monney, Claude and Jaouen, Thomas and Cario, Laurent and Calandra, Matteo and Cren, Tristan},
  title = {Misfit Layer Compounds: A Platform for Heavily Doped 2D Transition Metal Dichalcogenides},
  journal = {Advanced Functional Materials},
  volume = {31},
  pages = {2007706},
  year = {2021}
}

@article{Zullo,
  author = {Zullo, Ludovica and Marini, Giovanni and Cren, Tristan and Calandra, Matteo},
  title = {Misfit Layer Compounds as Ultratunable Field Effect Transistors: From Charge Transfer Control to Emergent Superconductivity},
  journal = {Nano Letters},
  volume = {23},
  number = {14},
  pages = {6658--6663},
  year = {2023}
}

@phdthesis{Zullo_thesis,
  author = {Zullo, Ludovica},
  title = {Emergent properties of misfit layer compounds from first principles},
  school = {Sorbonne Université and Università degli Studi di Trento},
  year = {2024}
}

@article{Zullo2,
  author = {Zullo, Ludovica and Setnikar, Grégory and Pawbake, Amit and Cren, Tristan and Brun, Christophe and Cordiez, Justine and Sasaki, Shunsuke and Cario, Laurent and Marini, Giovanni and Calandra, Matteo and Méasson, Marie-Aude},
  title = {Charge density wave collapse of NbSe2 in the (LaSe)1.14(NbSe2)2 misfit layer compound},
  journal = {Physical Review B},
  volume = {110},
  pages = {075430},
  year = {2024}
}

@article{Bianco_2020,
  title = {Weak Dimensionality Dependence and Dominant Role of Ionic Fluctuations in the Charge-Density-Wave Transition of ${\mathrm{NbSe}}_{2}$},
  author = {Bianco, Raffaello and Monacelli, Lorenzo and Calandra, Matteo and Mauri, Francesco and Errea, Ion},
  journal = {Physical Review Letters},
  volume = {125},
  issue = {10},
  pages = {106101},
  numpages = {6},
  year = {2020},
  month = {Sep},
  publisher = {American Physical Society},
  doi = {10.1103/PhysRevLett.125.106101},
  url = {https://link.aps.org/doi/10.1103/PhysRevLett.125.106101}
}

@article{Brumme2015,
  title = {First-Principles Theory of Field-Effect Doping in Transition-Metal Dichalcogenides: {{Structural}} Properties, Electronic Structure, {{Hall}} Coefficient, and Electrical Conductivity},
  shorttitle = {First-Principles Theory of Field-Effect Doping in Transition-Metal Dichalcogenides},
  author = {Brumme, Thomas and Calandra, Matteo and Mauri, Francesco},
  year = {2015},
  month = apr,
  journal = {Physical Review B},
  volume = {91},
  number = {15},
  pages = {155436},
  issn = {1098-0121, 1550-235X},
  doi = {10.1103/PhysRevB.91.155436}
}

@article{Sohier2017,
  title = {Density Functional Perturbation Theory for Gated Two-Dimensional Heterostructures: {{Theoretical}} Developments and Application to Flexural Phonons in Graphene},
  shorttitle = {Density Functional Perturbation Theory for Gated Two-Dimensional Heterostructures},
  author = {Sohier, Thibault and Calandra, Matteo and Mauri, Francesco},
  year = {2017},
  month = aug,
  journal = {Physical Review B},
  volume = {96},
  number = {7},
  pages = {075448},
  issn = {2469-9950, 2469-9969},
  doi = {10.1103/PhysRevB.96.075448}
  }

@article{MoS2_gated_1,
author = {J. M. Lu  and O. Zheliuk  and I. Leermakers  and N. F. Q. Yuan  and U. Zeitler  and K. T. Law  and J. T. Ye },
title = {Evidence for two-dimensional Ising superconductivity in gated MoS$_2$},
journal = {Science},
volume = {350},
number = {6266},
pages = {1353-1357},
year = {2015},
doi = {10.1126/science.aab2277}
}

@article{MoS2_gated_2,
author={Saito, Yu
and Nakamura, Yasuharu
and Bahramy, Mohammad Saeed
and Kohama, Yoshimitsu
and Ye, Jianting
and Kasahara, Yuichi
and Nakagawa, Yuji
and Onga, Masaru
and Tokunaga, Masashi
and Nojima, Tsutomu
and Yanase, Youichi
and Iwasa, Yoshihiro},
title={Superconductivity protected by spin--valley locking in ion-gated MoS$_2$},
journal={Nature Physics},
year={2016},
month={Feb},
day={01},
volume={12},
number={2},
pages={144-149},
issn={1745-2481},
doi={10.1038/nphys3580}
}

@article{Arguello2014,
  title = {Visualizing the charge density wave transition in $2H$-${\text{NbSe}}_{2}$ in real space},
  author = {Arguello, C. J. and Chockalingam, S. P. and Rosenthal, E. P. and Zhao, L. and Guti\'errez, C. and Kang, J. H. and Chung, W. C. and Fernandes, R. M. and Jia, S. and Millis, A. J. and Cava, R. J. and Pasupathy, A. N.},
  journal = {Physical Review B},
  volume = {89},
  issue = {23},
  pages = {235115},
  numpages = {9},
  year = {2014},
  month = {Jun},
  publisher = {American Physical Society},
  doi = {10.1103/PhysRevB.89.235115}
}

@article{weber2013,
  title = {Optical phonons and the soft mode in 2$H$-NbSe${}_{2}$},
  author = {Weber, F. and Hott, R. and Heid, R. and Bohnen, K.-P. and Rosenkranz, S. and Castellan, J.-P. and Osborn, R. and Said, A. H. and Leu, B. M. and Reznik, D.},
  journal = {Physical Review B},
  volume = {87},
  issue = {24},
  pages = {245111},
  numpages = {8},
  year = {2013},
  month = {Jun},
  publisher = {American Physical Society},
  doi = {10.1103/PhysRevB.87.245111},
}

@article{johannes_fermi_2008,
	title = {Fermi surface nesting and the origin of charge density waves in metals},
	volume = {77},
	copyright = {http://link.aps.org/licenses/aps-default-license},
	issn = {1098-0121, 1550-235X},
	url = {https://link.aps.org/doi/10.1103/PhysRevB.77.165135},
	doi = {10.1103/PhysRevB.77.165135},
	language = {en},
	number = {16},
	journal = {Physical Review B},
	author = {Johannes, M. D. and Mazin, I. I.},
	month = apr,
	year = {2008},
	pages = {165135},
}

@misc{sajan2026moireinduced,
      title={Moiré-induced symmetry breaking of charge order in van der Waals heterostructures}, 
      author={Sandra Sajan and Laura Pätzold and Tarushi Agarwal and Clara Pfister and Haojie Guo and Sisheng Duan and P. V. Sruthibhai and Mariana Rossi and Maria N. Gastiasoro and Sara Barja and Ravi P. Singh and Tim Wehling and Miguel M. Ugeda},
      year={2026},
      eprint={2603.05759},
      archivePrefix={arXiv},
      primaryClass={cond-mat.mes-hall},
      url={https://arxiv.org/abs/2603.05759}, 
}

@article{gao2018,
    author = {Shang Gao  and Felix Flicker  and Raman Sankar  and He Zhao  and Zheng Ren  and Bryan Rachmilowitz  and Sidhika Balachandar  and Fangcheng Chou  and Kenneth S. Burch  and Ziqiang Wang  and Jasper van Wezel  and Ilija Zeljkovic },
    title = {Atomic-scale strain manipulation of a charge density wave},
    journal = {Proceedings of the National Academy of Sciences},
    volume = {115},
    number = {27},
    pages = {6986-6990},
    year = {2018},
    doi = {10.1073/pnas.1718931115},
}

@article{chiu2025,
    author = {Chiu, Wei-Chi and Mardanya, Sougata and Markiewicz, Robert and Nieminen, Jouko and Singh, Bahadur and Hakioglu, Tugrul and Agarwal, Amit and Chang, Tay-Rong and Lin, Hsin and Bansil, Arun},
    title = {Strain-Induced Charge Density Waves with Emergent Topological  States in Monolayer NbSe2},
    journal = {ACS Nano},
    volume = {19},
    number = {19},
    pages = {18108-18116},
    year = {2025},
    doi = {10.1021/acsnano.4c13478},
}

@article{merino2025,
    author = {Luque Merino, Rafael and Carrascoso, Felix and Henríquez-Guerra, Eudomar and Calvo, M. Reyes and Frisenda, Riccardo and Castellanos-Gomez, Andres},
    title = {Strain Engineering of Correlated Charge-Ordered Phases in 1T-TaS2},
    journal = {Nano Letters},
    volume = {25},
    number = {46},
    pages = {16379-16386},
    year = {2025},
    doi = {10.1021/acs.nanolett.5c04101},
}

@article{zhang2017,
  title = {Strain engineering a $4a\ifmmode\times\else\texttimes\fi{}\ensuremath{\surd}3a$ charge-density-wave phase in transition-metal dichalcogenide $1\mathrm{T}\text{\ensuremath{-}}\mathrm{VS}{\mathrm{e}}_{2}$},
  author = {Zhang, Duming and Ha, Jeonghoon and Baek, Hongwoo and Chan, Yang-Hao and Natterer, Fabian D. and Myers, Alline F. and Schumacher, Joshua D. and Cullen, William G. and Davydov, Albert V. and Kuk, Young and Chou, M. Y. and Zhitenev, Nikolai B. and Stroscio, Joseph A.},
  journal = {Physical Review Mater.},
  volume = {1},
  issue = {2},
  pages = {024005},
  numpages = {10},
  year = {2017},
  month = {Jul},
  publisher = {American Physical Society},
  doi = {10.1103/PhysRevMaterials.1.024005}
}

@misc{le_du_2026,
      title={Tuning the Charge Transfer of Transition Metal Dichalcogenides via Misfit Layer Compounds}, 
      author={Le Du, Hugo and Ludovica Zullo and Justine Cordiez and Robin Salvatore and Giovanni Marini and Marie Hervé and Debora Pierucci and Shunsuke Sasaki and Florent Pawula and Etienne Janod and Chiara Bigi and Marta Zonno and François Bertran and Thomas Jaouen and Patrick Le Fèvre and Matteo Calandra and Laurent Cario and Tristan Cren and Marie D Angelo},
      year={2026},
      eprint={2605.18317},
      archivePrefix={arXiv},
      primaryClass={cond-mat.mtrl-sci},
      url={https://arxiv.org/abs/2605.18317}, 
}

@article{kundu_2024,
	title = {Charge density waves and the effects of uniaxial strain on the electronic structure of {2H}-{NbSe2}},
	volume = {5},
	issn = {2662-4443},
	url = {https://doi.org/10.1038/s43246-024-00661-7},
	doi = {10.1038/s43246-024-00661-7},
	number = {1},
	journal = {Communications Materials},
	author = {Kundu, Asish K. and Rajapitamahuni, Anil and Vescovo, Elio and Klimovskikh, Ilya I. and Berger, Helmuth and Valla, Tonica},
	month = oct,
	year = {2024},
	pages = {208},
}

@article{Monacelli_2021,
doi = {10.1088/1361-648X/ac066b},
url = {https://doi.org/10.1088/1361-648X/ac066b},
year = {2021},
month = {jul},
publisher = {IOP Publishing},
volume = {33},
number = {36},
pages = {363001},
author = {Monacelli, Lorenzo and Bianco, Raffaello and Cherubini, Marco and Calandra, Matteo and Errea, Ion and Mauri, Francesco},
title = {The stochastic self-consistent harmonic approximation: calculating vibrational properties of materials with full quantum and anharmonic effects},
journal = {Journal of Physics: Condensed Matter}
}
